\documentclass[sigconf,nonacm,screen]{acmart}
\usepackage[utf8]{inputenc}
\usepackage{booktabs}
\usepackage{tabularx}
\usepackage{multirow}
\usepackage{ifthen}
\usepackage{framed}
\usepackage{syntax}
\usepackage{comment}
\usepackage{float}
\usepackage{caption}
\usepackage{enumitem}
\usepackage{multicol}
\usepackage{ulem}
\usepackage{syntax}
\usepackage{framed}
\usepackage{amsmath,amsfonts}
\usepackage{amsthm}
\usepackage{array}
\usepackage{algorithmic}
\usepackage[ruled, vlined, linesnumbered]{algorithm2e}
\usepackage{color, colortbl}
\usepackage{listings}
\usepackage{soul}
\usepackage{graphicx}
\usepackage{dblfloatfix}
\usepackage{array}
\usepackage{balance}
\newcolumntype{?}{!{\vrule width 1pt}}
\newcolumntype{H}{>{\setbox0=\hbox\bgroup}c<{\egroup}@{}}

\graphicspath{ {images/} }

\newboolean{showcomments}
\setboolean{showcomments}{true} 

\ifthenelse{\boolean{showcomments}}{
    \newcommand{\nbc}[3]{
        {\colorbox{#3}{\bfseries\sffamily\scriptsize\textcolor{white}{#1}}}
        {\textcolor{#3}{\sf\small$\langle$\textit{#2}$\rangle$}}}
}{
    \newcommand{\nbc}[3]{}
    
}

\ifthenelse{\boolean{showcomments}}
{
    \newcommand{\del}[1]{\textcolor{red}{\sout{#1}}}    
}{
    \newcommand{\del}[1]{}                              
    
}

\newcommand\su[1]{\nbc{SU}{#1}{olive}}

\newcommand\vb[1]{\nbc{VB}{#1}{blue}}

\newcommand\resolved[1]{}

\makeatletter
\newcounter{examplealgorithm}

\makeatother

\newlist{myitemize}{itemize}{1}
\setlist[myitemize,1]{label=\textbullet,leftmargin=4mm}

\begin{document}
\title{Focused Dynamic Slicing for Large Applications using an Abstract Memory-Model}

\author{Alexis Soifer, Diego Garbervetsky, Victor Braberman}
\email{{asoifer,diegog,vbraber}@dc.uba.ar}
\affiliation{%
\institution{UBA, ICC, CONICET} \city{Buenos Aires}
\country{Argentina}}
\author{Sebastian Uchitel}
\email{suchitel@dc.uba.ar}
\affiliation{%
\institution{UBA, ICC, CONICET} \city{Buenos Aires}
\country{Argentina}}
\affiliation{%
\institution{Imperial College} \city{London} \country{UK}}


\renewcommand{\algorithmcfname}{Algorithm}
\SetKwInOut{Input}{Input}
\SetKwInOut{Output}{Output}
\SetKwFor{While}{while}{do}{}
\SetKwFor{ForEach}{for each}{do}{}
\SetKwFor{For}{for}{do}{}
\SetKwIF{If}{ElseIf}{Else}{if}{do}{else if}{else}{}
\SetKw{Return}{return}
\SetKw{Break}{break}
\SetKwProg{Fn}{Function}{ is}{end}
\SetKwProg{Def}{def}{:}{}

\newcommand{\node}{\textrm{n}}
\newcommand{\nodeABS}{\textrm{n}_\textrm{a}}
\newcommand{\nodes}{\textrm{N}}
\newcommand{\nodesState}{\textrm{N}_\textrm{S}}
\newcommand{\nodesStateB}{\textrm{N}_\textrm{S'}}
\newcommand{\nodesAbs}{\textrm{N}_\textrm{A}}
\newcommand{\edges}{\textrm{E}}
\newcommand{\edgesState}{\textrm{E}_\textrm{S}}
\newcommand{\edgesStateB}{\textrm{E}_\textrm{S'}}
\newcommand{\edgesAbs}{\textrm{E}_\textrm{A}}
\newcommand{\stateA}{\textrm{S}}
\newcommand{\stateB}{\textrm{S'}}
\newcommand{\abstractionA}{\textrm{A}}
\newcommand{\abstractionB}{\textrm{A'}}
\newcommand{\stateAbs}{\textrm{S}_\textrm{A}}
\newcommand{\sigmaS}{\sigma_\textrm{S}}
\newcommand{\sigmaSB}{\sigma_\textrm{S'}}
\newcommand{\sigmaAbs}{\sigma_\textrm{A}}
\newcommand{\sigmaAbsB}{\sigma_\textrm{A'}}
\newcommand{\wn}{\textrm{W}_\textrm{N}}
\newcommand{\wv}{\textrm{W}_\textrm{V}}
\newcommand{\rn}{\textrm{R}_\textrm{N}}
\newcommand{\rv}{\rn}
\newcommand{\types}{\overline{\textrm{Type}}}
\newcommand{\nodesSet}{2^{\textrm{N}}}
\newcommand{\nodesSetAbs}{2^{\textrm{N}_\textrm{A}}}
\newcommand{\lastDefSet}{2^{\textrm{Last Definition}}}

\newcommand{\nodesDG}{\textrm{N}}
\newcommand{\edgesDG}{\textrm{E}}
\newcommand{\reachableStmtDG}{\textrm{R}}
\newcommand{\stmtToNodesDG}{\textrm{S}}

\setlength{\grammarparsep}{0.05cm}

\definecolor{armygreen}{rgb}{0.29, 0.33, 0.13}
\lstset{
tabsize = 2, 
showstringspaces = false, 
numbers = left, 
commentstyle = \color{armygreen}, 
keywordstyle = \color{blue}, 
stringstyle = \color{red}, 
rulecolor = \color{black}, 
basicstyle = \small \ttfamily , 
breaklines = true, 
numberstyle = \tiny,
otherkeywords={*,var},
xleftmargin=1em,
framexleftmargin=1em,
language=[Sharp]C
}
\renewcommand{\lstlistingname}{Example}

\newcommand\todo[1]{\textcolor{blue}{#1}}

\begin{abstract}

Dynamic slicing techniques compute program dependencies to find all statements that affect the value of a variable at a program point for a specific execution.
Despite their many potential uses, applicability is limited by the fact that they typically cannot scale beyond small-sized applications.
We believe that at the heart of this limitation is the use of memory references to identify data-dependencies.
Particularly, working with memory references hinders distinct treatment of the code-to-be-sliced (e.g., classes the user has an interest in) from the rest of the code (including libraries and frameworks).
The ability to perform a coarser-grained analysis for the code that is not under focus may provide performance gains and could become one avenue toward scalability.

In this paper, we propose a novel approach that completely replaces memory reference registering and processing with a memory analysis model that works with program symbols (i.e., terms). In fact, this approach enables the alternative of not instrumenting -thus, not generating any trace- for code that is not part of the code-to-be-sliced.
We report on an implementation of an abstract dynamic slicer for C\#, \textit{DynAbs}, and  an evaluation that shows how large and relevant parts of Roslyn and Powershell -
two of the largest and modern C\# applications that can be found in GitHub- can be sliced for their test cases assertions in at most a few minutes. We also show how reducing the code-to-be-sliced focus can bring important speedups with marginal relative precision loss.


\end{abstract}

%


\maketitle

\section{Introduction}

Dynamic slicing ~\cite{Korel:1988:DPS:56378.56386} shows developers
which program statements impact the value of a variable at a specific line of code 
when the program is executed with a particular input. 
The study of dynamic slicing dates back over thirty years and its results have been 
applied to, for example, debugging and fault localization, cohesion measurement, program 
comprehension and testing \cite{10.1145/2544137.2544152, 9159078, 9610714, 8530064, 10.5555/2747476.2747526, DBLP:journals/corr/abs-1108-1352}.

\noindent
Dynamic slicing techniques compute program dependencies analyzing the 
information within the execution trace generated
through program or run-time environment instrumentation. 
Control and data dependencies of the trace statements are analyzed to
identifying those that affect the slicing criterion 
~\cite{Zhang:2005:EEU:1085130.1085135, Wang:2008:DSJ:1330017.1330021, hammacher-iwmse-2009}. 
Even though there are many papers that propose and optimize
dynamic slicing techniques, including trace compression, only a few dynamic slicing tools are publicly available and none are reported to scale to long executions of complex modern systems.

We believe one of the main obstacles to scalability of dynamic slicers is their need to record, trace, store, and later process concrete {\it memory references} (e.g., \cite{Zhang:2003:PDS:776816.776855}). Ultimately, those references are needed to query and get -from the slicer's internal data structures- the statement occurrence that last defined a given memory location. However, this introduces some difficulties: there is a time overhead on the instrumented code to get the memory references, trace size grows with the unavoidable need for compression (and decompression) \cite{Wang:2004:UCB:998675.999455}, this impacts size of internal data structures and time-cost of traversing them to get the last definitions \cite{Zhang:2003:PDS:776816.776855}.


 Working at the level of concrete memory references also hinders the ability to focus just on the  \textit{code-to-be-sliced}, which could be responsible for only a tiny fraction of the system trace. Of course, any part of the executed code may alter the def-use relationship between statements that are in the code-to-be-sliced, so in principle  memory operations of executed statements of, for instance, the library code (i.e., code that is not to be sliced) must be accounted for too, but this incurs in a performance penalty. Introducing within the memory-references paradigm a way to decouple library code is far from trivial and,
 to the best of our knowledge, 
 there are no general approaches that soundly handle arbitrary library code which may involve complex memory operations that can alias current and new allocated objects.
%
We believe the ability to define and focus on code-to-be-sliced is a must if one expects to handle modern and large framework-based applications in which most behavior is performed by either framework code or even application modules that are not relevant to the user.

We propose \textit{abstract dynamic slicing} that deals with accesses of memory locations through high-level program symbols (i.e., terms) instead of memory references. 
In contrast to traditional dynamic slicing, trace information 
collected only includes just the necessary data to identify executed statements, avoiding storage of concrete memory references accessed. 
This technique is based on a \textit{memory analysis model} (or just analysis model) that can map terms (access paths) into abstract memory locations. 
As slicing progresses over a program trace 
with no concrete memory location information, 
the memory model is updated and a dependency graph constructed. 

This approach naturally {supports diverse treatment of code}: when a call to a method which is not relevant for the user (or client analysis of the slicer) is {made, the code need not be instrumented nor its trace collected, simply} a conservative effect can be applied to the memory model -- i.e.,  a \textit{havoc} effect.
This consists in representing every possible read, write, or object allocation 
that could have been produced within the external method execution. The same can be done if the method relevant for the user is call-backed from methods that are not.
{Indeed, this allows soundly trading precision for scalability.}

In order to show that abstract dynamic slicing can be used for dealing
with modern and large applications we implemented \textit{DynAbs}, 
an abstract dynamic slicer for C\#.
We used DynAbs to slice some of the biggest C\# applications 
that can be found in GitHub, 
Roslyn -the C\# compiler- and Powershell -the Microsoft console using slicing criteria taken from tests found their GitHub projects.

The main contributions of this paper are: 
(1) a new sound technique called \textbf{abstract dynamic slicing} which that uses a  memory analysis model instead of real memory locations, (2) a tool, \textbf{DynAbs}, which implements 
the {abstract dynamic slicing} for C\# , and 
(3) an evaluation of DynAbs that includes $(i)$ a comparison with a state of the art tool on a common benchmark, and $(ii)$ an evaluation showing both that abstract dynamic slicing can handle  --even without trace compression-- slicing real test assertions in large C\# applications 
and how removing instrumentation can sometimes accelerate analysis with marginal loss of precision.



\section{Architecture and overview}
\label{sec:overview}
Our approach produces slices without having to trace all the executed code by treating symbolically the read/write effects of traced and untraced code.

\begin{figure}[htp]
\centering
\includegraphics[width=\columnwidth]{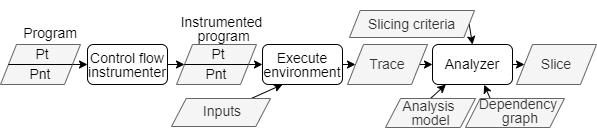}
\vspace{-4mm}
\caption{Abstract Dynamic Slicing Overview.}
\label{fig:architecture}
\end{figure}

Figure~\ref{fig:architecture} shows an overview of the approach.  We require an executable \textit{Program} separated into a portion that is to be traced ($P_T$) and another that is not ($P_{NT}$).
The former must be in an instrumentable format (i.e., in source code or bytecode), the
latter may include compiled third-party code\footnote{Note that $P_T$ should be a superset of the part of the code-to-be-sliced.}.
An \textit{Instrumenter} modifies the code-to-be-traced to record control
information. The code is executed with a user-chosen
\textit{Input} and a \textit{Trace} is produced that includes only indications of the blocks of code executed. Note that trace generation is lightweight, not recording concrete memory references, and thus,  avoiding instrumentation costs and the run-time overhead of recording every access to the memory.
The trace is then processed incrementally by the \textit{Analyzer} to construct a \textit{Dynamic Dependency Graph} (DDG) \cite{Agrawal}.
The trace is conceptually transformed into a stream of statements that is processed to update
both the DDG and a model that abstracts program's memory
(c.f., \textit{Memory Analysis Model}).
Finally, the analyzer outputs a program \textit{Slice} traversing the DDG from the
\textit{Slicing Criterion} provided by a user.


\noindent
This schema looks like classic dynamic slicing based on DDG (e.g., ~\cite{Agrawal}) but it solves the problem of determining the last definition of memory locations denoted by variables differently than later approaches and  tools (e.g., \cite{Wang:2004:UCB:998675.999455, Wang:2008:DSJ:1330017.1330021, hammacher-iwmse-2009, Azim:2019:DSA:3339505.3339649, 10.1145/3468264.3473123, Zhang:2004:WET:1038264.1038935}).
Our approach differs because we obtain symbolic terms (e.g, variable names, access paths) from the source code rather than actual memory references collected at run-time.

\begin{figure}[htp]
\centering
\includegraphics[width=\columnwidth]{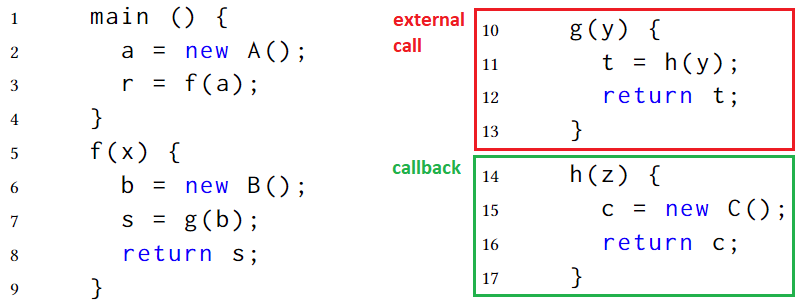}
\caption{Basic example. External call is not instrumented.}
\label{fig:ptgExampleCode}
\end{figure}

Consider the program in ~\autoref{fig:ptgExampleCode} that is to be sliced at line 3 (variable r).
A traditional slicing approach would instrument the whole program in order to produce a trace that registers the updates of each memory location accessed by the program.
For instance, upon execution of line 2 a tuple $\langle $2$^0, 0x8ff3, 0xefe9\rangle$
would be recorded in a trace file to indicate that in the first execution of line 2 the memory location of variable $a$ ($0x8ff3$) has the new memory allocated address $0xefe9$.
Later, a dependency graph would be updated using this information by first adding a node $n$ for that line for registering that the last update of $a$ is, effectively, at line $2^0$.

%

In contrast, in our approach a program execution trace stores the executed statements plus any
entrance and exit to control blocks. No memory references are stored. For the example above, the trace would be:
\begin{small}
\texttt{EnterMain$^0$, S2$^0$, S3$^0$, EnterF$^0$, S6$^0$,
S7$^0$, EnterH$^0$, S15$^0$, S16$^0$, ExitH$^0$, S8$^0$, ExitF$^0$, ExitMain$^0$}.
\end{small}


Informally, in our approach the processing of the trace proceeds as follows: when processing \texttt{2$^0$} our slicing algorithm tells the analysis memory model that a new object should be added.
After that, on reading \texttt{EnterF$^0$} it tells the analysis model to add a new context for \texttt{F}, and with \texttt{S6$^0$} that an object was created.
Something different occurs when processing \texttt{S7$^0$}.
Here, instead of processing a sub-trace corresponding to \texttt{g} (as in traditional approaches to slicing), the trace analyzer tells the analysis model  that a non-instrumented method (i.e., code in $P_{NT}$) was executed. The model conservatively updates the part of the head reachable from $b$ (the \textit{havoc} effect, see section~\ref{sec:underlyingModel}).
The trace analyzer can detect a non-instrumented call because
instead of receiving \texttt{EnterG$^0$}, it received \texttt{EnterH$^0$}.
This implies something else: a callback method has to be processed.
Hence, a conservative operation in the memory model assigns the method parameters (in this case, $z$), processes the sub-trace corresponding to in \texttt{h}, and conservatively assigns its returned value to the state of the non-instrumented method \texttt{g}.
Finally, the trace analyzer detects the exit from \texttt{g} when observing the statement \texttt{S8$^0$} from \texttt{f}.

Notice that the slice can work without the need of analyzing the whole program. In particular, it can avoid  tracing and analyzing a framework of library code that can represent thousand of even millions of line of code.
The price to pay is precision in the resulting slice.
In a traditional approach, lines 3, 7, 8, 11, 12, 15, and 16 will be part of the slice.
In contrast, this approach adds the line 6 although it has not been used.
In addition, lines 11 and 12 are not  part of the slice (as they are part of external, untraced and unsliced code).

In the next sections we introduce the analysis model we use to allow reasoning abstractly about the dependencies of a program execution trace.

\section{Points-to configuration Transition System}

We first provide a simple formalization of a (concrete) points-to configuration that comprises both the heap and the stack of a program. This allows formalizing how our memory analysis-model (presented in the next section) relates to actual memory snapshots of the program under analysis. 

\textbf{\textit{Terms}} are sequences of program symbols (and, potentially, concrete values for indexes in the case or arrays)  used to access memory locations.
The analysis model  works with \textbf{\textit{Static Terms}} or just \textbf{\textit{Terms}} (when it is clear from the context), which are sequences of program symbols (i.e., access paths with no concrete values for indexes). 
In what follows, consider that $last(t)$ returns the last field name of the term $t$, $base(t)$ returns the same term $t$ excluding only its last field. When $t$ is a variable name $v$, $last(v)=v$, $base(v)=\epsilon$. 
For the sake of simplifying presentations we omit dealing with types of terms and objects (see Section \ref{sec:implementation} for some details on how they are used to improve precision). Also, and without loss of generality, we avoid modeling scalar variables (but their treatment is very similar to variables that map to memory references). (Reference) variables are modelled as fields of potentially stacked environment objects.
\paragraph{Concrete Points-to Configuration}
Let $c ::= \langle O , E, \sigma \rangle$ be a concrete {\it points-to configuration} such that
$O \subseteq O_c$ where  $O_c$ is the universe of nodes representing dynamically 
allocated objects.  
$\edges \subseteq O \times \textrm{Fields} \times O$ (which is a partial function on the two first parameters)
is a set of labeled edges that represents values of object fields (i.e., $(n, f, m) \in 
\edges$ means that the field $f$ of object $n$ points to the object $m$), and 
$\sigma = \sigma_0 \dots \sigma_k$ is a stack of objects $\sigma_i\in O$, that 
stand for the current activation stack.  
Object $\sigma_0$ is the current activation object; its fields play the role of local 
variables and current method parameters.


\paragraph{Points-to Configuration Transitions} 
We now define the actions that make configurations transition to other configurations. Basically, actions are associated to the following basic operations: new-object (fresh object) allocation ($alloc (term)$), field assignment ($assign (term,term)$). 
Besides these basic Points-to operations, we need to take into account the effects of 
entering a method, which creates a new environment on top of stack with parameters pointing-to arguments,  ($enterMethod$ $(m,args,params)$), and exiting a method, that  pops the environment and assigns the return value to the corresponding left hand side field of the call to the method that has finalized ($exitMethod (m, lhs)$). 
Thus, the semantics is that of the standard points-to effects of those operations. We will denote such configuration transitions as $c \xrightarrow[]{op (p)} c'$ where $p$ is the operation's parameters. 

\paragraph{Configuration and Action Traces}. 
A Configuration Trace is  a sequence of consecutive configuration transitions, schematically, $c_0 \xrightarrow[]{op_1 (p_1)} c_1$ $\ldots$ $c_{n-1} \xrightarrow[]{op_n (p_n)} c_n$ generated by the program under analysis. For the sake of simplicity, lets assume, that $c_0$ is the empty configuration and the first operation, $op_1$, is $enterMethod(main,\_,\_)$ and stands for entering the $main$. We will call $CT(P)$ the set of all possible configurations traces of $P$.

\paragraph{Observable Configuration-Traces}
Now, Let $M2S$ (\textit{methods to be sliced}) be a subset of program methods that are of interest (i.e., the methods in the code-to-be-sliced). 
 Given a valid points-to trace $c_0 \xrightarrow[]{op_1 (p_1)} c_1....c_{n-1} \xrightarrow[]{op_n (p_n)} c_n$ we are interested in formally describing what the slicer can actually observe and infer about control flow and actions given that methods not in $M2S$ do not externalize actions. The observable trace is obtained by hiding (i.e., $\lambda$ transtion) all basic points-to actions executed by  methods not in $M2C$.  Entrances to methods will be then divided into four different categories and relabelled (and/or hidden) accordingly:
 \begin{itemize}
     \item  $enter++Method (m,params,args)$ when a method in $M2S$ is entered from a $M2S$'s method call-site, 
     \item $enter+-Method (m,args) $ when a method not in $M2S$ is entered from a $M2S$'s method call-site (e.g., a call to a non-instrumented library method), 
    \item $enter-+Method (m,params)$ when a method in $M2S$ is entered from a method that is not in $M2S$ (e.g., a call back from a non-instrumented part of a framework),
    \item $enter--Method ()$ when the a method not in $M2S$ is indicated to enter from a method not in $M2S$, those entrances are actually hidden. 
\end{itemize}

\begin{figure*}[t]
\centering
\includegraphics[width=500px]{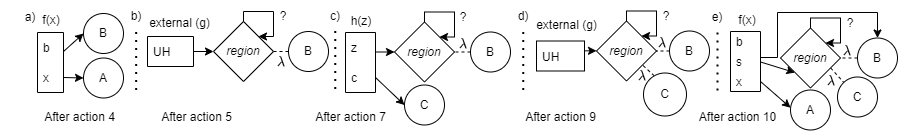}
\vspace{-2mm}
\caption{Snapshots of the memory analysis model during the trace consumption.}
\label{fig:ptgExample}
\vspace{-3mm}
\end{figure*}

Something analogous is applied to $exitMethod$ actions:  

\begin{itemize}
    \item $exit++Method (m,lhs)$ when a method in $M2S$ exits to a method in $M2S$ and the result is stored into $lhs$,
    \item $exit+-Method (m)$ when call-backed method-to-be-sliced exits with results in $result$  stored in an unobservable left hand side term, 
    \item $exit-+Method (m,lhs)$ when a method not in  $M2S$ exits to an instrumented method and the result is stored into $lhs$,
    \item $exit--Method (m)$ when the a method not in $M2S$ is indicated to exit into a method not in $M2S$, those exits are actually hidden.  
 \end{itemize}
Let $oCT(M2S)$ the set of configuration traces.
Even if the configuration of those traces are not actually observable by the slicer, their projection onto action label sequences can be inferred by the algorithm. For structured programs, those {\it observable action traces} are generated by the following grammar and, as shown later, that grammar provides the basis for getting the observable action trace from a stream of actions (some of them hidden).

\begin{framed}
\begin{grammar}

<start> ::= <++Method>  | <NoStatements> 

<++Method> ::= enter++Method  <Statements> exit++Method

<Statements> ::=  <basicPTaction> <Statements> |
<++Method> <Statements>
|
<+-Method> <Statements> | $\epsilon$

<+-Method> ::= enter+-Method  <NoStatements> exit-+Method

<NoStatements> ::=  <-+Method> <NoStatements> | $\epsilon$

<-+Method> :: = enter-+Method <Statements> exit+-Method

<basicPTaction> ::= alloc | assign
\end{grammar}
\end{framed}

On the other hand, the set of configuration traces also defines a new transition system, \textit{Observable Points-to Configuration LTS}, for the program $P$ with the same configurations and new transitions for the new labels as follows: $enter++, exit++, enter-+, exit +-$ are defined as the transitions Points-to Configuration LTS. For the case of $enter+-$ and $exit-+$, transitions relates the source configuration to any configuration reachable by following 0 or more hidden actions.

Following the initial example in \autoref{fig:ptgExampleCode}, and by defining $M2S = \{\texttt{main}, \texttt{f}, \texttt{h} \}$ the observable action trace associated to the control-flow trace presented in Section~\ref{sec:overview} is as follows: 
\begin{small}
\texttt {1. enter++Method$_{main}$}, 
\texttt{2. alloc(a)},
\texttt{3. enter++Method$_{f}$(x,a)}, 
\texttt{4. alloc(b)}, 
\texttt{5. enter+-Method$_{g}$(b)},\\
\texttt{6. enter-+Method$_{h}$(z)},
\texttt{7. alloc(c)}, 
\texttt{8. assign(rv,c)}, 
\texttt{9. exit+-Method$_{h}$}(), \\
\texttt{10. exit-+Method$_{g}$}(s),
\texttt{11. assign(rv,s)},
\texttt{12. exit++Method$_{f}$}(r), 
\texttt{13. exit++Method$_{main}$.}
\end{small}

\vspace{-4mm}
\section{Memory Analysis Model}
\label{sec:underlyingModel}

\begin{table*}[t]
\vspace{-4mm}
\footnotesize
\begin{center}
\begin{tabular}{|p{30mm}|p{140mm}|}
\hline
Operation & Postcondition \\
\hline
$m$.init$() =m'$ & $m'$ is an empty analysis model  
\\
\hline	
 $m$.get$(t :: \texttt{Term})=O$ & $O$ is the set of all abstract objects
 reached by traversing the transitions in $\hat{E}$  by following the sequence of fields in $t$ starting at $\sigma_0$ via matching transitions (or wildcard transitions) and taking $\lambda$ related objects at each navigation step. Note that $m$.get$(\epsilon) = \hat{\sigma_0}$. 
\\
\hline
$m$.alloc$(t :: \texttt{Term}, n:: DDG node)=m'$ & 
$m'$ is like $m$ but now for all abstract objects $b \in m.$get$(base(t))$, $o' \in m'.$get$(b.last(t))$, where $o'$ is a new singleton denoting the newly allocated object. 
$n$ is added to the set $lastdef [last(t)]$ of $lhs$ objects. If $b$ happens to be just one analysis object marked as singleton whatever $lastdef[last(t)]$ mapped to is replaced by  $\{n\}$ and $o'= m.$get$(b.last(t))$ (strong update). \\

\hline
$m$.assign$(t_i, t_j :: \texttt{Term}, n :: DDG node)=m'$ & 
$m'$ is like $m$ but now for all abstract objects $b \in m.$get$(base(t_i))$, $t \in  m.$get$(t_j),
t \in m'.$get$(b.last(t_i))$.
If $m.$get$(b)$ happens to be just one analysis object marked as singleton then the new edges replaces the old ones (strong update). Something analogous happens with $n$ added to $b.lastdef[last(t)]$.
\\
\hline
$m$.enter++Method$(\vec{p}, \vec{a} :: \overline{\texttt{Term}}, $n$:: DDG node) = m'$ & 
Pushes new local variables environment into the stack ($\sigma'$ denotes the new stack in $m'$) and assigns arguments $\vec{a}$ to formal parameters names $\vec{p}$ of the pushed environment object (i.e, $m'$.get$(\vec{p}[i]) = $m$.get(\vec{a}[i])$ for all $i$). 
It records $n$ (the call-site) as the last definition node for formal parameters $p$.
\\
\hline
m.exit++Method$(lhs :: \texttt{Term}, n :: DDG node)=m'$ &  
Pops $m$'s current environment $\sigma_0$ and assigns $\sigma'_0.lhs$ ($\sigma'$ denotes the new stack in $m'$) to a distinguished environment field $\sigma_0.rv$ from the previous environment. 
This assignment follows the same logic for determining if a strong update is possible. \\
\hline
$m$.enter+-Method$(\Bar{a}::\overline{\texttt{Term}}, n :: DDG node)=m'$  &  
Let $m(\Bar{a})$ all the reachable objs. in $m$ from those terms. 
$m'$ adds to $m$ a new region $r$ (with a $?$-annotated self-loop transition). 
$r$ is $\lambda$-equivalent to every object in  $m(\Bar{a})$. 
Also, $r.LastDef[?] = \{n\}$ and, $m'$ has a new environment on top in such that $m'.\sigma_0.UpdatedHeap$ = $r$. \\
\hline
$m$.exit-+Method$(lhs:: \texttt{Term}, n :: DDG node)=m'$ & 
Pops current environment $\sigma_0$ and assigns $\sigma'_0.lhs$ to the special field $\sigma_0.UpdatedHeap$ from the previous environment.\\
\hline
$m$.enter-+Method$(\Bar{p}::\overline{\texttt{Term}}, n :: DDG node)=m'$ & 
Pushes a new environment $\sigma_0$ featuring formal parameters $p$ of the call-backed method that points to the caller region, i.e.,  $m'$.get$(p) = m.$get$(UpdatedHeap)$ using previous environment.\\
\hline
$m$.exit+-Method$(n :: DDG node)=m' $ &  
Pops the current environment and $m'$.get$(UpdatedHeap)$ is $\lambda$-equivalent to $m$.get$(rv)$ using previous environment.
Also, $\forall o \in$ m'.get($UpdatedHeap$) $: \ n \in o.LastDef[?]$. \\
\hline
\end{tabular}
\caption{Analysis Model Interface}
\label{tab:modelInterface}
\end{center}
\vspace{-8mm}
\end{table*}

The key point of abstract dynamic slicing is that it works symbolically with 
the effects of read/writes in an execution. 
By symbolic,  we mean that the analysis of the trace works with syntactic terms instead of memory references. More concretely, a data dependency of DDG between node $r$ and $w$ is added when $w$ stands for a statement occurrence that may have be the last update of one of the underlying memory locations that is denoted by a (syntactic) term $t$ occurring in the statement associated to node $r$. Thus, the core functionality of the analysis model is to $get$ the potential (last) writers of a memory location that could be currently dereferenced by term $t$.



\textbf{\textit{Analysis Model Configuration}}.
Like the points-to configuration, a configuration of the analysis model is, in essence, a graph. However, nodes may represent more than one concrete object of the points-to state (e.g, all objects of an array, all objects created by a method, etc.) and fields (labeled edges) might point to more than one abstract object to handle the fact that one may not be able to precisely determine where terms point to when symbolically tracking memory updates. That is, let $m ::= \langle \hat{O}, \hat{O_s},\lambda, \hat{E}, \hat{\sigma}
\rangle$ be a configuration of the 
analysis model where $\hat{O} \subseteq O_a$ is a set of abstract objects, 
some of them $\hat{O}_s\subseteq \hat{O}$ are marked as singletons (the others are called regions),
$\hat{E}  \subseteq \hat{O} \times \textrm{ Field} \cup \{ ? \} \times \hat{O}$
represents the connections between abstract objects. 
Connections may be via  field names or a special wildcard symbol \textit{?}, which 
underspecifies the field responsible for the connection. 
$\lambda \subseteq \hat{O} \times \hat{O}$ is an equivalence relation between model objects used to indicate that those model objects may be denoting the same concrete object.
Sequence  $\hat{\sigma}$ denotes abstract stacks of abstract activation objects.


Identifying singleton abstract nodes enables a more precise treatment of updates as 
will be indicated later on. 
A key observer operation of the Analysis Model is $get$ which given a term $t$ 
 returns the set of all abstract objects 
 that are reached by 
 traversing the transitions in $\hat{E}$ 
 by following the sequence of fields in $t$ starting at $\sigma_0$ via matching 
 transitions (or wildcard transitions) and taking $\lambda$ related objects at each navigation step. Note that $get(\epsilon) = \hat{\sigma_0}$.


\textit{\textbf{Sound Abstraction}}. 
To define which concrete points-to configurations are 
modelled  by an analysis model configuration we use a denotation function $\delta$.
Given $c ::= \langle O , E, \sigma
\rangle$ a concrete configuration, 
$m ::= \langle \hat{O}, \hat{O_s}, 
\hat{E}, \lambda, \hat{\sigma}
\rangle$ a configuration of the analysis model, 
and a function $\delta:  \hat{O} 
\mapsto 2^{O}$, we say that $m$ soundly abstracts $c$ according to $\delta$ 
(noted as $abs(m,c,\delta)$) if the following hold:

\begin{enumerate}

\item $\forall \hat{o} \in \hat{O}_s.|\delta(\hat{o})|=1$
\item $\forall \hat{o}_1, \hat{o}_2 \in \hat{O}, \delta(\hat{o}_1) \cap 
\delta(\hat{o}_2) \neq \emptyset \implies (\hat{o}_1, \hat{o}_2)\in \lambda$
\item $\forall o \in O, \exists \hat{o}  \in \hat{O} , o \in \delta(\hat{o}) $
\item $\forall \hat{o} \in \hat{O}, \forall o \in   \delta(\hat{o}), \forall o' \in O, o 
\xrightarrow{f} o'  \implies \exists \hat{o}', \hat{o}'', \hat{o}''' \in \hat{O}. (o' \in  \delta(\hat{o}')) \land 
(\hat{o}'' \xrightarrow{f} \hat{o}'''$ or $\hat{o}'' \xrightarrow{?} \hat{o}''') \land (\hat{o}, \hat{o}'') \in \lambda \land (\hat{o}''', \hat{o}') \in \lambda $ 
\item  $|\sigma|=|\hat{\sigma}|=k$ 
and $\wedge \, \forall i=0..k,$ 
$\{ \sigma_i\}=\delta(\hat{\sigma}_i)$

\end{enumerate}

The first property says singleton objects map to exactly one object. The second and 
third, that concrete objects are abstracted by exactly one 
$\lambda$-partition of abstract objects. The fourth property states that navigation through fields is preserved in partitions
(although a wildcard field can be used). The fifth, that the denotation function should 
maintain the correspondence between concrete and abstract local variable 
environments. 
Finally, the type of abstract object should be compatible 
with those of denoted objects.
 Given two denotation functions $\delta$ and $\delta'$, we say that $\delta'$ extends $\delta$ ($\delta \subseteq  \delta'$) iff $dom(\delta) \subseteq  
dom(\delta')$ and $\forall a \in dom(\delta), (\delta(a) \subseteq \delta'(a))$.

\textbf{\textit{Analysis-Model Update}}. 
The slicing algorithm we present maintains a current analysis model configuration that, after processing an action occurrence, is an abstract version of the  concrete points-to configuration reached by the application at that point (which is not explicitly informed to the algorithm).
Given the actions on the observable concrete points-to configurations LTS presented in the previous section, analysis model features the corresponding actions ($m 
\xrightarrow[]{\hat{op}} m'$ 
and they abstractly mimic the evolution of  the concrete points-to configuration by just knowing the observable actions with (static) terms. This is done because, while processing the stream of statements of the trace, a key invariant is that there exists a denotation function between the   the analysis model and the current (and unknown in detail) concrete points-to configuration\footnote{ This also implies simulation and thus the existence of access paths mentioned by terms in the analysis model.}. Thus, we design these corresponding operations  to  extend denotation functions. 
More formally, the memory model ``API'' must guarantee that for any operation $op$ and its abstract counterpart $\hat{op}$\footnote{
 As traces do not contain concrete execution values, the slicing algorithm has no  
knowledge of array index values (i.e., does not handle dynamic terms). Thus, the (static) term is actually a virtual projection of the actual term (trimming index values), meaning array accesses need to be treated as if they had been done to an indeterminate index.}, for all concrete configurations $c$, $c'$, denotation 
function $\delta$ and analysis-model configuration $m$ we have: 

\noindent $ abs(m,c,\delta) \ \wedge \ c 
\xrightarrow[]{op} c'$ implies 
$\exists \delta'
\wedge \ 
\delta \subseteq \delta' \land 
abs(\hat{op} (m),c',\delta')$.


%
%

Beyond keeping the abstract shape of points-to configuration, the analysis model is also used by the algorithm to store (and later query) information about which nodes of the DDG \textit{may} be the last definitions for those denoted memory locations. More precisely, a model object yielded by the analysis model contains a dictionary that maps field names to set of nodes of the DDG (e.g., $\hat{o}.lastDef[fie$ $ld]$, or similarly $\hat{\sigma}_0. lastdef[var]$) with $\hat{o}$ and  $\hat{\sigma}_0$ being an abstract object and an abstract root (current environment) object respectively). 
The invariant kept by the algorithm over this analysis model guarantees that, given a term $t$, $\bigcup_{\hat{o}\in get(base(t))} \hat{o}.lastDef[last(t)]$ includes all the last definition nodes for the last field of the term $t$. 
In table ~\ref{tab:modelInterface}, the corresponding operations of the analysis model are explained. Their postconditions are meant to preserve the correspondence with the concrete points-to configuration as defined above. 

The enter and exit method actions that cross the boundaries of what is informed by the infrastructure (i.e., $enter+-$, $enter-+$, $exit+-$, $exit-+$) are designed to conservatively deal with several challenges arising due to non-observation of one or more actions that are not included in the trace (by {\it havoc}king an area of the heap model). 
For instance, $enter+-$ must match the configuration without knowing neither the formal parameters nor the sequence of effects done by the actions that follow entering the method not in $M2S$. This is done by assuming conservatively that all what is reachable from the arguments is subject to read and write \footnote{Global variables accessible by the $M2C$ should also be added to be sound.}. For the case of $exit-+$ model does not know what actually is being returned and, thus, it escapes the region standing for all potentially altered objects through $lhs$. In $enter-+$, arguments are unknown and thus parameters are matched to whatever the current non-instrumented method might have in its reachable part of the heap. For $exit+-$,  although it is known where the return value is stored, what is not known is where non-instrumented method is going to store or use that part of the heap. Thus, return value  is added to its current reachable heap.   
From the control-flow trace of \autoref{sec:overview}, the algorithm presented in next section would apply a sequence of analysis model updates that map one-to-one with the  observable action trace of \autoref{sec:underlyingModel}. 
In ~\autoref{fig:ptgExample} we illustrate the snapshots some points of that sequence of operations (see captions).

\section{Abstract Dynamic Slicing Algorithm}

\def\HiLi{\leavevmode\rlap{\hbox to \hsize{\color{yellow!50}\leaders\hrule height 
.8\baselineskip depth .5ex\hfill}}}

\setlength{\textfloatsep}{0pt}
\setlength{\floatsep}{0pt}
\begin{algorithm}[tbh]
  \Input{P :: \texttt{Program}, M2S:: \texttt{Set[MethodNames]}, i :: \texttt{Input},  c :: \texttt{Criterion}}
  \Output{\texttt{Slice}}
  \SetNoFillComment
  \texttt{global} $ddg \leftarrow \textrm{\textit{create dynamic dependency graph}}$ 
  \\
  \texttt{global} $model \leftarrow \textrm{\textit{create analysis model}}$ \\
   \texttt{global} $statActionStream \leftarrow StatActionStream(execute(instrument(p,M2S), i)))$
   $s \leftarrow peek(statActionStream)$ \texttt{\small // look ahead}\\
  
  \Switch{$s.actionType$}
	{
		\Case{$enter++Method$}{
			$Process++Method()$ \\
	}
		\Case{$enter-+Method$}{
			$ProcessNoStatements$ \\
	}
	}
  \Return{ddg.slice(\textrm{\textbf{c}})}
  \caption{Start Process}
\end{algorithm}

\begin{algorithm}[htb]
   \SetNoFillComment
   $s \leftarrow consume(statActionStream)$ \texttt{\small // It must be an enter++Method (those are generated by the filter pre-processing) and carry all necessary information about parameters and arguments}\\
   $n \leftarrow UpdateDDG(s, s.args.uses)$  \texttt{\small // (potentially) new node into the DDG that deppends on last def of args}\\
   $model.enter++Method(s.args,s.params,n)$ \\ \texttt{\small // model is notified}\\
   $ProcessStatements()$ \\
   $s \leftarrow consume(statActionStream)$ \texttt{\small // It must be a return statement that will be treated as an exit++Method}\\
   $l \leftarrow consume(statActionStream)$ \texttt{\small // Thus, this must be the lhs assignment}\\
   $n' \leftarrow UpdateDDG(l, rv)$ \texttt{\small // ``rv'' is a distinguished identifier the model use for the return value}\\ 
   $model.exit++Method(l.lhs,n')$ \texttt{\small // update the model to reflect it has exited }\\
   \Return{}
    \caption{Process++Method}
\end{algorithm}

\begin{algorithm}[h]
   \SetNoFillComment
   $s \leftarrow consume(statActionStream)$ \texttt{\small // It must be an enter+-Method}\\
   $n \leftarrow UpdateDDG(s, s.args.uses)$ \\ 
   $model.enter+-Method(s.args, n)$ \\
   $ProcessNoStatements()$ \\  
   $s \leftarrow consume(statActionStream)$ \texttt{\small //a statement different from method enter (no call back) is at the peek of the stream. This indicates the flow has returned to the last instrumented part that called the processed un-instrumented method}\\
   $l \leftarrow consume(statActionStream)$ \texttt{\small // Thus, this is the lhs assignment for the return value of the the method not in M2C (i.e. uninstrumented)}\\
   $n' \leftarrow UpdateDDGwithNode(l, n)$ \texttt{\small $n$ is the DDG node standing for the call to the +-Method   // }\\ 
   $model.exit-+Method(l.lhs)$ \texttt{\small // update the model to reflect it has exited }\\
   \Return{}
    \caption{Process+-Method}
\end{algorithm}

\begin{algorithm}[h]
   \SetNoFillComment
   $s \leftarrow consume(statActionStream)$ \texttt{\small // It must be an enter method}\\
   $n \leftarrow UpdateDDG(s, s.args.uses)$ \\ 
   $model.enter-+Method(s.params, n)$ \\
   $ProcessStatements()$ \\  
   $s \leftarrow consume(statActionStream)$ \texttt{\small this is the return from the call back}\\
   $model.exit+-Method()$ \\
   \Return{}
    \caption{Process-+Method}
\end{algorithm}

\begin{algorithm}[t!b!h!] 
  \SetNoFillComment
  $s \leftarrow peek(statActionStream)$\\
  \While{$! s.type=return)$}
  {
	
	\Switch{$s.actionType$}
	{
	  
		\Case{$\textrm{assignment}$}{    
				\label{line:assignStart}
				$n \leftarrow UpdateDDG(s, s.uses)$ \\ \label{line:getFreshNode}
\label{line:assignEndClassic}
					$model.assign(s.lhs, s.rhs, n)$ \\ \label{line:assignEnd}
					$s \leftarrow \textrm{\textit{consume  statActionStream}}$ \\
	}
	
 \Case{$\textrm{object creation}$} { 
				\label{line:ObjCreateStart}
					$n \leftarrow UpdateDDG(s, \emptyset)$ \\
					$model.alloc(s.lhs, n)$ 
					\\ \label{line:alloc} \label{line:ObjCreateEnd}
					$s \leftarrow \textrm{\textit{consume  statActionStream}}$ \\
	}

  \Case{$\textrm{enter++Method}$ }{
					Process++Method() \\ 
	}
 \Case{$\textrm{enter+-Method}$ }{
					Process+-Method() \\
	}

	}	
	}

\Return{}
  
  \caption{$ProcessStatements$}
\end{algorithm}

\setlength{\textfloatsep}{0pt}
\begin{algorithm}[t!b!h!] 
  \SetNoFillComment
  $s \leftarrow peek(statActionStream)$\\
  \While{($s.type=MethodEnter$)}
  {
					Process-+Method() \texttt{\small // call-back detected}\\
	
	}	
	
\Return{}
  
  \caption{$ProcessNoStatements$}
\end{algorithm}

\begin{algorithm}[tb]
  \Input{s :: \texttt{Statement},  uses :: $Set[\texttt{Term}]$}
  \SetNoFillComment
  \label{line:getControlNode}
  $dd \leftarrow \emptyset$ \\
  \ForEach{$t \in uses$} {
	\label{line:getLastDefsStart}
	$objs \leftarrow model.get(base(t))$ \\
	$dd \leftarrow dd \cup \{\;lastDef[o][last(t)]\;|\;o \in objs\;\}$ \\
	\label{line:getLastDefsEnd}
  }
  $n \leftarrow ddg.addNode(s, \{ cd \} )$\\
  \label{line:addDependency}
  \Return{$n$}
  \caption{Update DDG}
  \label{alg:saveDeps}
\end{algorithm}

In Algorithms 1 through 7 
we show the abstract slicing algorithm. It works a grammar directed translation to transduce the visible points-to actions into analysis model actions. It adds a couple of concerns as well: it identifies and categorizes some of the entrance and exit to methods to define which grammar production to process, it also updates the DDG.
We omit the treatment of control dependencies as it is an orthogonal concern to the technique. The implementation uses a stack of DDG nodes associated to control statements to know always the last relevant control action that explains why a statement is being executed. 

Algorithm 1 corresponds to the initial production. It initializes the main data structures. Particularly, it builds up the stream of statements and actions that is subject to the parsing and transducing phase later on. More concretely, $StatActionStream$ is the function that yields a stream of statements modified to feature also some of the categories of enter/exit method actions underlying the control flow followed by the application. 
$StatActionStream$ can be regarded as the composition of two filters: one uses compilation infrastructure to convert the control-flow trace yielded by (partial) instrumentation into a stream of statements. Those statements include compilation information like parseable access path that play the role of static terms. 
Statements, by demand, yield which terms are being read and which term is being defined. We also assume that invocation statements and statements that stand for the entry point of a method can be queried for the name and number of arguments of the method being called or, respectively, entered ($s.method$).

The other filter, has as purpose to replace some of the enter/return method statements with higher-level actions using the following logic: when an invocation statement $s$ is followed by a method enter $l$ such that $s.method$ = $l.method$ then an enter++Method with arguments and parameters replace both statements. When, $s.method$ and $l.method$ are different then enter+-Method with the invocation arguments replaces the invocation statement \footnote{If the statement stream starts with $s$ such that $s.method$ = $main$ an enter++Method is added at the front of the stream, enter-+Method, otherwise}. 

The other possible combinations are not detected by this filter and are actually detected by the algorithm while parsing the statement action stream by following the grammar presented previously.
enter-+Method: is the case during $NoStatements$ processing when method enter occurs in the trace (i.e, an enter without a call has happened), exit++Method during $Statements$ processing when a return appears and $Statements$ was executed from the context of a $++Method$ processing,   exit+-Method when a return statement arrives while processing Statements in the context of processing a -+Method, and exit-+Method, during $NoStatements$ processing when a statement different from method enter occurs in the trace.

Method $UpdateDDG$ deals with dependencies. As mentioned, the approach handles data-dependencies using the analysis model. 
In lines 2-4 of Algorithm 7, abstract
objects are obtained from
the analysis model, and are used together with the last field of each term to collect data dependencies.
Finally (line 5), adds a node to the dependency graph 
with the collected dependencies.

\section{Implementation}
\label{sec:implementation}

We developed a tool, DynAbs, that implements the abstract dynamic slicing algorithm presented above for core features of the C\# language. 
We discuss in this section some key design decisions aiming at performance.
%

%

{\textbf{\textit{Dynamic Program Dependency Graph}}}.
We implement a graph where nodes represent statement occurrences and edges represent 
data and control dependencies between executed statements. 
To avoid a linear growth of this structure we follow the solution proposed in 
~\cite{Agrawal} that only creates a new node when a statement 
occurrence does not share the same dependencies as a previous occurrence (a subsumption of transitive dependencies).
%

\textbf{\textit{{Analysis Model}}}
It is worth mentioning that as $\lambda$ partitions the set of abstract nodes,
we use Disjoint-Set Union data structure \cite{10.1145/321879.321884} to summarize and quickly access information in partition representatives that would be spread all over the  nodes of the partition otherwise (rendering imprecision also a factor for being slower \cite{10.1145/3236454.3236500}). 
Besides, when created, objects are actually annotated with type information  using what the compilation infrastructure knows about statements in the program. Then, for  model operations that receive terms as input and need to find all possible model objects that need to be queried or modified, the type information in the terms passed as parameters is used to filter out objects whose annotated type is not compatible. In fact, partition representatives stores information qualified by types.

\section{Evaluation}
\label{sec:evaluation}

Now, we describe our experimental setup and discuss the
evaluation results. The general aim of the evaluation is to assess if DynAbs can slice large applications with reasonable precision and performance. Given that existing slicing tools cannot process such applications and that manual slicing of large applications is extremely challenging, error prone and laborious, we do not perform an an evaluation of precision in absolute terms, i.e., against a ground truth. Instead, we compare precision against the state of the art under conditions in which this comparison is possible, and then evaluate relative precision loss in large applications. In summary, to evaluate DynAbs, we answer the following research questions:


\noindent
\textbf{RQ1:} {To what extent is DynAbs precise when compared to the state-of-the-art in dynamic slicing?}\\
\textbf{RQ2:} {Can DynAbs slice large apps. using real slicing criteria?}\\
\textbf{RQ3:} {To what extent does DynAbs lose precision (and gain performance) when non-traced code is increased in large applications}?

\subsection{Experimental Setup}


To answer RQ1, we compared DynAbs with the  
state-of-the-art dynamic slicer, Javaslicer \cite{hammacher-iwmse-2009}. 
Note that the more recent Java slicer, Slicer4J, 
supports the latest Java features that Javaslicer does not but has worst performance \cite{10.1145/3468264.3473123}. 
Also note that there are no publicly available slicers for C\#.

To compare a Java and a C\# Slicer, we require a benchmark with programs that can be written in syntactically (and sematically) equivalent versions for both languages.
We chose Olden \cite{Carlisle95, 10.1145/209936.209941} in its Java version, JOlden \cite{10.5555/645988.674177}, 
and we manually wrote its version for C\# 
preserving line per line, the syntactic structure of the programs. 
Key in the choice of Olden is that programs do not contain calls to library, thus, translation effort is contained and controlled, and slices can be completely analyzed.
Note that, olden has been used as a benchmark in numerous studies
\cite{10.1145/2647508.2647524, 10.1145/2660193.2660194, NAVARRO200747, 10.1145/844102.844112}.
JOlden consists of a collection of 10 programs that includes small and medium-sized scientific code (bh and em3d), process simulations (health and power), graph optimization routines (mst and tsp), graphics utilities (perimeter and voronoi), a sorting routine (bisort) and a toy tree benchmark (treeadd).

As slicing criteria, we selected the lines of code that contained variables storing the final
result values calculated during the executions.
If one of these results was an instance of a complex element (e.g., class, struct)
then we selected the instance and its fields.
In the case of arrays, we selected the array and its first element.
To select program inputs, consider the fact that the Olden programs take as input the size of the problem to be solved (e.g., number of cities for the travelling salesperson problem) and then randomly pick an instance of that size. We selected minimal sized inputs to facilitate comparing slice sizes, and fixed a random instance with that size to allow reproducibility. 

To answer RQ2 and RQ3, we chose 
Roslyn \cite{Roslyn} and Powershell \cite{Powershell}.
Roslyn, the .NET Compiler Platform, 
is the biggest C\# application on GitHub.
It is a set of open-source compilers and code analysis 
APIs for .NET languages. 
On the other hand, Powershell, 
the Microsoft console,
is the highest ranked C\# application 
with more than 500K LoC.
To identify realistic slicing scenarios we chose as slicing criteria assertions in existing test cases and selected the classes under test as the code-to-be-sliced, \textit{havoc}-ing the rest of the code. We compared precision and performance of computing these slices against slicing for the same criteria but with an \textit{extended instrumentation} of the code.

More specifically, for Roslyn we did the following. The core of compilers project in Roslyn (\texttt{roslyn/src/Compilers}) consists of three main folders. The first contains generic implementations and base classes for every .NET language (\texttt{core}, 83K LoC). The other two are the \texttt{VisualBasic}  and \texttt{CSharp} specific implementations, 
with 206K LoC and 230K LoC respectively.  We chose, the largest of the these two (i.e., \texttt{.../Compilers/CSharp}). 

The test folder for the C\# compiler code (i.e., \texttt{.../CSharp/Test}) is loosely structured according to the roslyn official pipeline architecture \cite{Roslyn} 
which describes four APIs: Syntax Tree,  Symbols, Binder and Flow Analyzer, and Emit. We selected one relevant test class for each of the APIs except for the Binder and Flow Analyzer API for which we chose two test classes (one for the Binder and one for the Flow analysis). The classes selected were: StatementParsingTests.cs, TypeTests.cs, BindingTests.cs, FlowTests.cs, and CompilationEmitTests.cs. For each of the classes we randomly selected for each test class, 20 tests.

Establishing the code-to-be-sliced required determining what code might be relevant in a debugging scenario when a tests fails. We took a syntactic approach based on the names of source code folders for this. We selected from the \texttt{.../CSharp/Portable} folder any subfolder that has a name that corresponds to the API under test. For instance, for the syntax tree API tests taken from StatementParsingTests.cs, we instrumented both the tests and all the classes in the  \texttt{.../Portable/Syntax} folder. 
In \autoref{tab:tablePSRoslyn}, column \textit{Inst. Code} indicates the percentage of instrumented code for each test class with respect to the total code in \texttt{.../Portable}.
We then computed slices for every assertion in every test. 
In addition, as an extension to the previous instrumentation we instrumented all of the code in \texttt{.../Portable} and generated the slices for the same criteria as before to allow comparing differences in terms of performance and precision. 


In  PowerShell, tests can be of two kinds: based on external scripts or using xUnit test framework. Only the latter are written in C\# and are amenable to slicing using DynAbs. We used all the tests in \texttt{./}\texttt{test/}\texttt{xUnit} \texttt{/csharp} which, as documented, contains all the C\# tests for the PowerShell Core project. To select the code-to-be-sliced we inspected each test (both code and assertions) to determine the classes it referenced directly.  
We then generated slices for each assertion in each test.
To produce the additional set of slices with an extended instrumentation of the code, we selected for instrumentation folder \texttt{./src/System.Management.Automation} which subsumes all of the previously instrumented classes. An important note is that both Roslyn and Powershell code include some advanced C\# language features that the DynAbs instrumentation does not support. To handle these cases, manual intervention of the instrumented code was required in a few specific locations.

\subsection{Results}

\textbf{RQ 1.} \autoref{tab:tableOlden} shows for each subject, in the first three columns, the number of different slicing criteria that were used and the average size of executions (measured in statements) and also the average number of unique statements (or lines) that were covered in the executions. The table also shows for DynAbs and Javaslicer the number of statements in the resulting slice, the relative size of the slice and the time to compute it. 

In BiSort, Perimeter, TreeAdd, and TSP, there are no array structures (nor calls to external libraries). This means that, by construction, DynAbs slices are sound and precise, assuming no implementation bugs. 
However, in BiSort, Javaslicer produces a slice that is smaller 
but unsound due to previously reported \cite{10.1145/3468264.3473123} unsound treatment of recursion and return statements dependencies.


The remaining 6 subjects include arrays which are a source of imprecision for DynAbs. However, DynAbs slices were smaller in 4 of them (BH, Em3d, Health, and Voronoi). Javaslicer's additional imprecision in these 4 subjects is related to its inaccurate modelling of control dependencies.
Finally, in MST and Power,  DynAbs imprecision with arrays leads to a 14\% and 5\% increment in slice size. Note that although DynAbs shows improved performance over Javaslicer, such comparison should not be considered relevant as the choice of slice tasks with minimal sized inputs is biased towards DynAbs (smaller inputs means smaller execution sizes). Indeed, note that in Power, which has a significantly larger execution size than the rest, trace compression of Javaslicer starts to show better performance than DynAbs. 

In summary, only in 3 out of 10 cases did DynAbs produce larger slices than Javaslicer. One of these (BiSort) is due to unsoundness of Javaslicer. Consecuently, only in 2 out of 10 cases were DynAbs slices less precise than Javalicer.\\


\begin{table}[thb]
    \vspace{-5mm}
	\small
	\setlength\tabcolsep{3.3pt}
	\begin{tabular}
		{|p{10mm}|p{2mm}|p{6mm}|p{3mm}|p{3mm}|p{2mm}|p{5mm}|p{3mm}|p{2mm}|p{5mm}|}
		\hline
		\multicolumn{4}{|l|}{ } & 
		\multicolumn{3}{|c|}{DynAbs} & 
		\multicolumn{3}{|c|}{Javaslicer} \\
		\hline
		Subject & 
		\multicolumn{1}{|l|}{Criteria} &		
		ES &
		\multicolumn{1}{|l|}{Stms.} & 
		SS &
		SS (\%) &
		t (s) &
		SS &
		SS (\%) &
		t (s) \\
		\hline
		BH & \multicolumn{1}{|r|}{6} & \multicolumn{1}{|r|}{3626} & \multicolumn{1}{|r|}{356} & \multicolumn{1}{|r|}{120} & \multicolumn{1}{|r|}{34} & \multicolumn{1}{|r|}{2.1} & \multicolumn{1}{|r|}{141} & \multicolumn{1}{|r|}{40} & \multicolumn{1}{|r|}{9.5} \\
        BiSort & \multicolumn{1}{|r|}{3} & \multicolumn{1}{|r|}{347} & \multicolumn{1}{|r|}{115} & \multicolumn{1}{|r|}{60} & \multicolumn{1}{|r|}{52} & \multicolumn{1}{|r|}{0.4} & \multicolumn{1}{|r|}{57} & \multicolumn{1}{|r|}{50} & \multicolumn{1}{|r|}{9.7} \\
        Em3d & \multicolumn{1}{|r|}{5} & \multicolumn{1}{|r|}{444} & \multicolumn{1}{|r|}{146} & \multicolumn{1}{|r|}{55} & \multicolumn{1}{|r|}{38} & \multicolumn{1}{|r|}{0.5} & \multicolumn{1}{|r|}{66} & \multicolumn{1}{|r|}{45} & \multicolumn{1}{|r|}{9.0} \\
        Health & \multicolumn{1}{|r|}{3} & \multicolumn{1}{|r|}{1377} & \multicolumn{1}{|r|}{187} & \multicolumn{1}{|r|}{29} & \multicolumn{1}{|r|}{16} & \multicolumn{1}{|r|}{0.7} & \multicolumn{1}{|r|}{36} & \multicolumn{1}{|r|}{19} & \multicolumn{1}{|r|}{9.0} \\
        MST & \multicolumn{1}{|r|}{1} & \multicolumn{1}{|r|}{802} & \multicolumn{1}{|r|}{187} & \multicolumn{1}{|r|}{111} & \multicolumn{1}{|r|}{59} & \multicolumn{1}{|r|}{0.4} & \multicolumn{1}{|r|}{97} & \multicolumn{1}{|r|}{52} & \multicolumn{1}{|r|}{8.0} \\
        Perimeter & \multicolumn{1}{|r|}{2} & \multicolumn{1}{|r|}{113} & \multicolumn{1}{|r|}{111} & \multicolumn{1}{|r|}{28} & \multicolumn{1}{|r|}{25} & \multicolumn{1}{|r|}{0.3} & \multicolumn{1}{|r|}{40} & \multicolumn{1}{|r|}{36} & \multicolumn{1}{|r|}{8.5} \\
        Power & \multicolumn{1}{|r|}{4} & \multicolumn{1}{|r|}{227496} & \multicolumn{1}{|r|}{296} & \multicolumn{1}{|r|}{201} & \multicolumn{1}{|r|}{68} & \multicolumn{1}{|r|}{39.3} & \multicolumn{1}{|r|}{192} & \multicolumn{1}{|r|}{65} & \multicolumn{1}{|r|}{10.8} \\
        TreeAdd & \multicolumn{1}{|r|}{1} & \multicolumn{1}{|r|}{59} & \multicolumn{1}{|r|}{44} & \multicolumn{1}{|r|}{20} & \multicolumn{1}{|r|}{45} & \multicolumn{1}{|r|}{0.3} & \multicolumn{1}{|r|}{20} & \multicolumn{1}{|r|}{45} & \multicolumn{1}{|r|}{9.0} \\
        TSP & \multicolumn{1}{|r|}{5} & \multicolumn{1}{|r|}{665} & \multicolumn{1}{|r|}{147} & \multicolumn{1}{|r|}{57} & \multicolumn{1}{|r|}{39} & \multicolumn{1}{|r|}{0.5} & \multicolumn{1}{|r|}{59} & \multicolumn{1}{|r|}{40} & \multicolumn{1}{|r|}{8.6} \\
        Voronoi & \multicolumn{1}{|r|}{6} & \multicolumn{1}{|r|}{1490} & \multicolumn{1}{|r|}{319} & \multicolumn{1}{|r|}{163} & \multicolumn{1}{|r|}{51} & \multicolumn{1}{|r|}{0.9} & \multicolumn{1}{|r|}{177} & \multicolumn{1}{|r|}{56} & \multicolumn{1}{|r|}{9.0} \\
		\hline
	\end{tabular}
	\caption{Olden results. $ES$: average execution size in \# of statements, $Stms.$: average number of unique statements that can be sliced, $SS$: average number of statements in the slice, $SS (\%)$:$SS/Stms.$, and $t$ is the average slice computation time in seconds.}
	\label{tab:tableOlden}
	\vspace{-7mm}
\end{table}

\begin{table*}[h]
	\small
	\setlength\tabcolsep{3.3pt}
	\begin{tabular}
		{|p{25mm}|p{8mm}|
			p{7mm}|p{7mm}|p{6mm}|p{5mm}|
			p{5mm}|p{8mm}|p{9mm}|p{15mm}|
			p{6mm}|p{2mm}|p{5mm}|}
		\hline
		\multicolumn{2}{|l|}{ } & 
		\multicolumn{8}{|c|}{Code-to-be-sliced instrumentation only} & 
		\multicolumn{3}{|c|}{Extended  Instrumentation} \\
		\hline
		Test class files (.cs)& 
		Criteria &		
		ES & 
		Stmts. & 
		SS & 
		SS(\%) & 
		t (s) &
		Inst.(\%) & 
		Speedup &
		Prec. loss(\%) &
		ES' & 
		SS' & 
		t (s) \\
		\hline
		\multicolumn{13}{|l|}{Roslyn} \\
		\hline
StatementParsingTests & \multicolumn{1}{|r|}{534} & \multicolumn{1}{|r|}{19270} & \multicolumn{1}{|r|}{1228} & \multicolumn{1}{|r|}{416} & \multicolumn{1}{|r|}{33.85} & \multicolumn{1}{|r|}{3.8} & \multicolumn{1}{|r|}{9.02} & \multicolumn{1}{|r|}{x1} & \multicolumn{1}{|r|}{0.00} & \multicolumn{1}{|r|}{19338} & \multicolumn{1}{|r|}{416} & \multicolumn{1}{|r|}{3.9} \\
TypeTests & \multicolumn{1}{|r|}{171} & \multicolumn{1}{|r|}{49211} & \multicolumn{1}{|r|}{3982} & \multicolumn{1}{|r|}{1586} & \multicolumn{1}{|r|}{39.84} & \multicolumn{1}{|r|}{23.9} & \multicolumn{1}{|r|}{35.85} & \multicolumn{1}{|r|}{x2} & \multicolumn{1}{|r|}{0.00} & \multicolumn{1}{|r|}{76391} & \multicolumn{1}{|r|}{1586} & \multicolumn{1}{|r|}{53.2} \\
BindingTests & \multicolumn{1}{|r|}{67} & \multicolumn{1}{|r|}{3176} & \multicolumn{1}{|r|}{974} & \multicolumn{1}{|r|}{258} & \multicolumn{1}{|r|}{26.45} & \multicolumn{1}{|r|}{1.1} & \multicolumn{1}{|r|}{14.41} & \multicolumn{1}{|r|}{x93} & \multicolumn{1}{|r|}{0.78} & \multicolumn{1}{|r|}{105219} & \multicolumn{1}{|r|}{256} & \multicolumn{1}{|r|}{101.7} \\
FlowTests & \multicolumn{1}{|r|}{40} & \multicolumn{1}{|r|}{667} & \multicolumn{1}{|r|}{388} & \multicolumn{1}{|r|}{104} & \multicolumn{1}{|r|}{26.76} & \multicolumn{1}{|r|}{0.4} & \multicolumn{1}{|r|}{4.53} & \multicolumn{1}{|r|}{x218} & \multicolumn{1}{|r|}{0.00} & \multicolumn{1}{|r|}{106715} & \multicolumn{1}{|r|}{104} & \multicolumn{1}{|r|}{82.6} \\
CompilationEmitTests & \multicolumn{1}{|r|}{99} & \multicolumn{1}{|r|}{3582} & \multicolumn{1}{|r|}{887} & \multicolumn{1}{|r|}{280} & \multicolumn{1}{|r|}{31.58} & \multicolumn{1}{|r|}{1.6} & \multicolumn{1}{|r|}{6.32} & \multicolumn{1}{|r|}{x81} & \multicolumn{1}{|r|}{0.00} & \multicolumn{1}{|r|}{105464} & \multicolumn{1}{|r|}{280} & \multicolumn{1}{|r|}{128.2} \\
		\hline
		\multicolumn{13}{|l|}{Powershell} \\
		\hline		
		FileSystemProvider & \multicolumn{1}{|r|}{13} & \multicolumn{1}{|r|}{14410} & \multicolumn{1}{|r|}{633} & \multicolumn{1}{|r|}{13} & \multicolumn{1}{|r|}{2.00} & \multicolumn{1}{|r|}{2.3} & \multicolumn{1}{|r|}{4.89} & \multicolumn{1}{|r|}{x91} & \multicolumn{1}{|r|}{0.00} & \multicolumn{1}{|r|}{151088} & \multicolumn{1}{|r|}{13} & \multicolumn{1}{|r|}{214.3} \\
MshSnapinInfo & \multicolumn{1}{|r|}{1} & \multicolumn{1}{|r|}{96} & \multicolumn{1}{|r|}{104} & \multicolumn{1}{|r|}{6} & \multicolumn{1}{|r|}{5.77} & \multicolumn{1}{|r|}{0.1} & \multicolumn{1}{|r|}{0.28} & \multicolumn{1}{|r|}{x1} & \multicolumn{1}{|r|}{0.00} & \multicolumn{1}{|r|}{557} & \multicolumn{1}{|r|}{6} & \multicolumn{1}{|r|}{0.1} \\
NamedPipe & \multicolumn{1}{|r|}{4} & \multicolumn{1}{|r|}{209} & \multicolumn{1}{|r|}{156} & \multicolumn{1}{|r|}{5} & \multicolumn{1}{|r|}{2.89} & \multicolumn{1}{|r|}{0.1} & \multicolumn{1}{|r|}{14.88} & \multicolumn{1}{|r|}{x1} & \multicolumn{1}{|r|}{0.00} & \multicolumn{1}{|r|}{299} & \multicolumn{1}{|r|}{5} & \multicolumn{1}{|r|}{0.1} \\
CorePsPlatform & \multicolumn{1}{|r|}{1} & \multicolumn{1}{|r|}{9} & \multicolumn{1}{|r|}{13} & \multicolumn{1}{|r|}{2} & \multicolumn{1}{|r|}{15.38} & \multicolumn{1}{|r|}{0.1} & \multicolumn{1}{|r|}{3.30} & \multicolumn{1}{|r|}{x1} & \multicolumn{1}{|r|}{0.00} & \multicolumn{1}{|r|}{9} & \multicolumn{1}{|r|}{2} & \multicolumn{1}{|r|}{0.1} \\
PowerShellAPI & \multicolumn{1}{|r|}{3} & \multicolumn{1}{|r|}{1728} & \multicolumn{1}{|r|}{1066} & \multicolumn{1}{|r|}{491} & \multicolumn{1}{|r|}{46.06} & \multicolumn{1}{|r|}{0.8} & \multicolumn{1}{|r|}{4.16} & \multicolumn{1}{|r|}{x981} & \multicolumn{1}{|r|}{0.00} & \multicolumn{1}{|r|}{225696} & \multicolumn{1}{|r|}{491} & \multicolumn{1}{|r|}{825.9} \\
PSConfiguration & \multicolumn{1}{|r|}{286} & \multicolumn{1}{|r|}{312} & \multicolumn{1}{|r|}{171} & \multicolumn{1}{|r|}{44} & \multicolumn{1}{|r|}{25.65} & \multicolumn{1}{|r|}{0.1} & \multicolumn{1}{|r|}{0.10} & \multicolumn{1}{|r|}{x1} & \multicolumn{1}{|r|}{7.32} & \multicolumn{1}{|r|}{642} & \multicolumn{1}{|r|}{41} & \multicolumn{1}{|r|}{0.1} \\
Binders & \multicolumn{1}{|r|}{1} & \multicolumn{1}{|r|}{12} & \multicolumn{1}{|r|}{18} & \multicolumn{1}{|r|}{4} & \multicolumn{1}{|r|}{22.22} & \multicolumn{1}{|r|}{0.1} & \multicolumn{1}{|r|}{4.00} & \multicolumn{1}{|r|}{x1} & \multicolumn{1}{|r|}{0.00} & \multicolumn{1}{|r|}{12} & \multicolumn{1}{|r|}{4} & \multicolumn{1}{|r|}{0.1} \\
PSObject & \multicolumn{1}{|r|}{15} & \multicolumn{1}{|r|}{496} & \multicolumn{1}{|r|}{299} & \multicolumn{1}{|r|}{78} & \multicolumn{1}{|r|}{25.97} & \multicolumn{1}{|r|}{0.1} & \multicolumn{1}{|r|}{1.19} & \multicolumn{1}{|r|}{x43} & \multicolumn{1}{|r|}{0.00} & \multicolumn{1}{|r|}{8456} & \multicolumn{1}{|r|}{78} & \multicolumn{1}{|r|}{4.3} \\
ExtensionMethods & \multicolumn{1}{|r|}{1} & \multicolumn{1}{|r|}{98} & \multicolumn{1}{|r|}{76} & \multicolumn{1}{|r|}{21} & \multicolumn{1}{|r|}{27.63} & \multicolumn{1}{|r|}{0.2} & \multicolumn{1}{|r|}{3.72} & \multicolumn{1}{|r|}{x6} & \multicolumn{1}{|r|}{600.00} & \multicolumn{1}{|r|}{3290} & \multicolumn{1}{|r|}{3} & \multicolumn{1}{|r|}{1.0} \\
PSVersionInfo & \multicolumn{1}{|r|}{1} & \multicolumn{1}{|r|}{202} & \multicolumn{1}{|r|}{149} & \multicolumn{1}{|r|}{11} & \multicolumn{1}{|r|}{7.38} & \multicolumn{1}{|r|}{0.1} & \multicolumn{1}{|r|}{0.17} & \multicolumn{1}{|r|}{x1} & \multicolumn{1}{|r|}{0.00} & \multicolumn{1}{|r|}{202} & \multicolumn{1}{|r|}{11} & \multicolumn{1}{|r|}{0.1} \\
Runspace & \multicolumn{1}{|r|}{9} & \multicolumn{1}{|r|}{2536} & \multicolumn{1}{|r|}{1130} & \multicolumn{1}{|r|}{548} & \multicolumn{1}{|r|}{48.48} & \multicolumn{1}{|r|}{0.9} & \multicolumn{1}{|r|}{8.37} & \multicolumn{1}{|r|}{x38} & \multicolumn{1}{|r|}{0.00} & \multicolumn{1}{|r|}{105198} & \multicolumn{1}{|r|}{548} & \multicolumn{1}{|r|}{34.1} \\
SecuritySupport & \multicolumn{1}{|r|}{1} & \multicolumn{1}{|r|}{41} & \multicolumn{1}{|r|}{48} & \multicolumn{1}{|r|}{20} & \multicolumn{1}{|r|}{41.67} & \multicolumn{1}{|r|}{0.1} & \multicolumn{1}{|r|}{1.33} & \multicolumn{1}{|r|}{x1} & \multicolumn{1}{|r|}{0.00} & \multicolumn{1}{|r|}{104} & \multicolumn{1}{|r|}{20} & \multicolumn{1}{|r|}{0.1} \\
SessionState & \multicolumn{1}{|r|}{1} & \multicolumn{1}{|r|}{159105} & \multicolumn{1}{|r|}{6217} & \multicolumn{1}{|r|}{4} & \multicolumn{1}{|r|}{0.06} & \multicolumn{1}{|r|}{165.2} & \multicolumn{1}{|r|}{67.98} & \multicolumn{1}{|r|}{x1} & \multicolumn{1}{|r|}{0.00} & \multicolumn{1}{|r|}{179189} & \multicolumn{1}{|r|}{4} & \multicolumn{1}{|r|}{219.9} \\
Utils & \multicolumn{1}{|r|}{45} & \multicolumn{1}{|r|}{20} & \multicolumn{1}{|r|}{9} & \multicolumn{1}{|r|}{5} & \multicolumn{1}{|r|}{53.60} & \multicolumn{1}{|r|}{0.1} & \multicolumn{1}{|r|}{0.40} & \multicolumn{1}{|r|}{x2} & \multicolumn{1}{|r|}{0.00} & \multicolumn{1}{|r|}{1619} & \multicolumn{1}{|r|}{5} & \multicolumn{1}{|r|}{0.2} \\
WildcardPattern & \multicolumn{1}{|r|}{13} & \multicolumn{1}{|r|}{6} & \multicolumn{1}{|r|}{10} & \multicolumn{1}{|r|}{2} & \multicolumn{1}{|r|}{16.91} & \multicolumn{1}{|r|}{0.1} & \multicolumn{1}{|r|}{0.22} & \multicolumn{1}{|r|}{x1} & \multicolumn{1}{|r|}{0.00} & \multicolumn{1}{|r|}{12} & \multicolumn{1}{|r|}{2} & \multicolumn{1}{|r|}{0.1} \\
		\hline
	\end{tabular}
	\caption{Results for Roslyn and Powershell. $Criteria$:number of slicing criteria used, $ES$: average execution size in \# of statements, 	$Stmts.$: average number of unique statements that can be sliced, $SS$ ($SS'$):average slice size in number of statements and $SS (\%)$ ($SS' (\%)$): $SS / Stmts.$ ($SS' / Stmts.$), $t$: average slice computation time in seconds, $Inst. (\%)$: percentage of instrumented compared to that instrumented in the "Extended Instrumentation", and $Speedup$ and $Prec. loss$: the slicing speedup time and precision loss percentage of the less instrumented code compared to the slicing of the code with extended instrumentation.}
	\label{tab:tablePSRoslyn}
	\vspace{-8mm}
\end{table*}

\noindent
\textbf{RQ 2.} 
The first columns of Table \ref{tab:tablePSRoslyn} show how DynAbs performed when slicing for assertions in real Roslyn and PowerShell tests. For each test class we show the number of criteria for which slices were produced, the average execution size (in number of statements) and the average number of unique statements (lines of code) that can be sliced. We  report on the size of the resulting slice and the reduction that the slice represents over the total number of unique statements. We also report the average slice computation time after compilation. This is because, due to the size of the projects, compilation is an order of magnitude larger (in the order or minutes) than the execution, tracing plus analysis time that we report. Finally, we report on the number of calls to external code, that is calls from the instrumented code to non-instrumented (and \textit{havoc}-ed) code. Recall that these external calls are not traced nor analyzed, and may represent complex behaviour that contains many more calls to non-instrumented code.

Results show that DynAbs took on average per slice less than a few seconds for all but two classes (TypeTests and SessionState). Within the various slices computed for each class there was little variation in performance. In fact, no slice for tests other than from TypeTests and SessionState took over than 5 seconds. Slices for TypeTests had an average of 24s, with a maximum time of 56s. The only slice for SessionState took 165s.
Slice size shows significant reduction in many cases (up to 98\%) although in some cases reduction was less than 50\%. 

In summary, DynAbs was able to slice most real tests for Roslyn and Powershell in a few seconds (over 1300 slices). Only few slices took (less than 2\%) up to 1 minute, and only one took 2.75 minutes. \\

\noindent
\textbf{RQ 3.} 
The rightmost columns of Table \ref{tab:tablePSRoslyn} show the results of slicing the same files as in RQ2 but with additional information extracted from an execution in which there is more instrumented source code. As before, we report the average execution size (in number of statements). We also report on the size of the slice and report the average slice computation time. As in RQ2, we report computation time after compilation.

Column \textit{Inst (\%)} shows the percentage of code that was initially instrumented (the code-to-be-sliced) with respect to the total code instrumented (and for which the rightmost columns report on). Note that for all test classes, the original instrumentation is less than 10\% of the extended instrumentation. In other words, the extended instrumentation adds a significant amount of additional code that is traced and analysed by DynAbs. The additional instrumentation can also be observed when comparing the columns for execution size. The execution sizes for the extended instrumentated code can grow over two orders of magnitud (e.g., FlowTests, CompilationEmitTests, PowerShellApi).

The \textit{Speedup} column in Table \ref{tab:tablePSRoslyn} shows the factor by which performance is improved in the original (less intrumented) slicing computations when compared to the slices computed with more instrumented code. As expected, classes in which execution size is originally significantly smaller have major speedups (x10 to x981). 

The \textit{Precision Loss} column captures, proportionally, how many statements were added to the slice computed with less instrumentation but that thanks to the additional instrumentation were actually not true depenedencies of the slice criteria. In other words, it is the average slice size for code-to-be-sliced instrumentation divided the average slice size for the extended instrumentation, minus 1.


Results show that for Roslyn, additional instrumentation does not result in increased precision. Indeed, the precision loss for only instrumenting the code-to-be-sliced (and no additional code, abstracted away using havocs) is on average marginal. Only 12 of the 171 slices computed for TypeTest lost precision and their worst case was 0.2\%, and only 54 slices computed for BindingTests lost precision and their worst case was 2\%. On the other hand average speedups by reduced instrumentation are achieved in 4 out of 5 case studies, 3 of which are very significant (x81 to x218). 

Speedup results for Roslyn show as expected that the less instrumented code based has significant speedups when the extended instrumentation makes execution sizes larger. In three 3 out of 5 case studies, the three with the largest increment of average execution sizes, the speedup was x81, x93 and x218.

For Powershell, additional instrumentation only led to improved slices in two of the test classes (PSConfiguration and ExtensionMethods). In other words, the precision loss for only instrumenting the code-to-be-sliced (and no additional code) is on average marginal for 13 out of 15 classes. PSConfiguration had an average precision loss of 7.32\%  and the only slice for ExtensionMethods a precision loss of 600\%. Zooming in on PSConfiguration, only 82 slices of the 286 had a precision loss, averaging 175\% and with a worst case of 400\%.

Speedup results for Powershell slices also show that less instrumented code based has significant speedups in particular when execution size increase in the more instrumented code base. For 8 out of 15 test classes, there was no speedup (nor precision loss). In these (e.g., NamedPipe) the execution size of the less instrumented code base is not significantly less than that of the more instrumented version, thus the lack of performance increase. However, there are 5 out of 15 test clases for which there is a speedup with no precision loss at all. PSVersionInfo, that has a significant precision loss also has an important speedup. PSConfiguration's has no average speedup but when looking only at the average speedup of the 82 slices that incurred in a 175\% precision loss, their average speedup is x2.9.

In summary, keeping instrumentation focused on the code-to-be-sliced of real Roslyn tests did not lose precision and achieved an average speedup of at least x2 in 4 out of 5 test class files, with significant increases for 3 out 5 (greater than x81). On the other hand, only 83 out of 395 slices taken from real Powershell tests exhibited significant precision losses when keeping instrumentation limited to the code-to-be-sliced, the rest had no precision loss whatsoever. Average speedups were achieved  in 40\% of the test classes and for those slices that had precision loss, performance gains were x2.9. 

These results provide evidence that keeping instrumentation and hence tracing focused to the code-to-be-sliced, and \textit{havoc}-ing external code can increase performance significantly without paying a proportional cost in terms of precision.

\subsection{Limitations and Threats to Validity}
Our experimentation is subject to various threats to validity. The focus on only two large case studies is a threat to generalization. For Powershell we used all available C\# tests, however for Roslyn manual selection was required as number of available tests was very large. We believe we mitigated bias by ensuring that tests classes covered the different architectural components and selected test classes based on the relevance of their names with respect to the name of the architectural component they belonged to. Where possibly the largest threat lies is in the distinction that we made between the code-to-be-sliced and the rest. That is, we manually selected source code files that a user when slicing for a test assertion may wish to have sliced, and selected a superset of more indirect dependencies as external code that is not to be sliced but that can be instrumented. This distinction, although made systematically to avoid bias, may draw the line differently from what a real user may want. However, we believe that our selection is reasonable and aims to demonstrate the tradeoffs that are possible when augmenting instrumented code without augmenting the code-to-be-sliced. 






\section{Related Work}
\label{sec:relatedWork}

Dynamic slicing (e.g.,\cite{Agrawal,Korel:1988:DPS:56378.56386}) has been regarded as a potentially useful idea but hard to scale ~\cite{Binkley:2014:OLP:2635868.2635893,Gupta:1997:HSI:261640.261644,Zhanglongrunning,Zhang:2004:CED:996841.996855,Wang:2004:UCB:998675.999455,10.1145/1075382.1075384,articleCDPSR,6693066,10.1109/ICSE-Companion.2019.00136}. For instance, due to the scalability issues, works like \cite{10.1007/s10664-020-09931-7} end up resorting to filtering out static slices with information collected at run-time on code actually executed. The use of a static slice might render such sort of approaches too imprecise, for instance, when dealing with large classes hierarchies and rich inputs.

Besides compression~\cite{Wang:2004:UCB:998675.999455,Wang:2008:DSJ:1330017.1330021,hammacher-iwmse-2009} there exist a few other approaches to deal with size.
\cite{Zhanglongrunning}  works with user-delimited part of the execution. Even approaches based on record \& replay technology like ~\cite{10.1145/2544137.2544152}, that collect memory references at replay time, still require users to keep small ``buggy regions'' to avoid incurring in large run-time overheads.

In statistical program slicing
~\cite{10.1109/ICSE-Companion.2019.00136} 
control and data-flow information is selectively 
collected  employing a hardware extension
in order to  compute an \textit{observed} program dependency graph.
This graph is formed by nodes which represent static instructions 
and arcs represent control or data dependencies in any of the monitored executions.
This approach deals with the scalability problem of tracing runtime at the cost of reducing accuracy (i.e., missing some statements). 
ORBS~\cite{Binkley:2014:OLP:2635868.2635893}  is based on a different notion of slice (observational), thus it avoids the instrumentation of the program.
The cost of this approach is in the need of multiple compilation and executions. 
Precision is sacrificed when insufficient iterations are performed.
It is worth mentioning ~\cite{10.1145/3468264.3473123} in which, based on taint analysis summaries, library code is replaced with assignments that produce the same memory effects as the code being removed. Note that, this enables basic def/use modeling but it does not, in general, neither handle potential aliasing effects nor it is meant to be a mechanism able to avoid instrumentation of arbitrary methods.
Independently and similarly to us, authors of Mandoline tool ~\cite{mandoline} follows an approach that do not resort to dynamic tracking object references (for slicing Android apps).Unlike our approach, Mandoline identifies variables referencing the same object via alias analysis over the trace of statements. Given a use of a field, several backward and forward traversals of portion of the trace till the set of all reaching definitions is built. Code that is out of scope of slicing must be either replaced by summaries in the form of assignments or by a built-in modeling of data propagation inside Android's framework methods. 

The static analysis community has addressed the abstract representation of heaps for different client analyses ~\cite{10.1145/2931098}. 
Our analysis model is ``storeless''~\cite{10.1145/2931098} in the sense that is meant to answer questions about access paths (in fact, it can ask if two scope-valid access paths may alias). However, its formulation is ``store-based'' since we establish an easy denotation relationship with the concrete points-to model, plus we want to operate with a graph view point when doing summarization. In fact, the context of use of our analysis model, dynamic analysis, implies our need of summarization is not iteration or recursion but the access to indexed structures and the need for accounting for effects of uninstrumented methods. Summarization of access paths is done in a graph-based fashion by adding regions and establishing $\lambda$-equivalence among nodes.

\section{Conclusion and future work}
The contribution of this paper is a sound dynamic slicing approach, Abstract Dynamic Slicing, 
that just need to generate a control-flow trace of methods-to-be-sliced. It then updates and uses an abstract memory model to query about how the memory may be configured. 
The approach enables users to tradeoff between precision and performance by setting the code that is instrumented. 
We report an implementation of the approach, \textit{DynAbs}, for C\# and experimentation that provides evidence that Abstract Dynamic Slicing can slice test case assertions of large C\# applications. 

Future work may include defining summaries to avoid always applying the most conservative model updates for common external libraries and systematically dealing with multi-threading.

\end{document}